\documentclass[amsmath,amssymb]{epl}
\usepackage{wasysym}% symbols

\title{Parity-dependent Kondo effect in ultrasmall metallic grains} 
\author{Giancarlo Franzese\inst{1,2}, Roberto Raimondi\inst{3}
 \and Rosario Fazio\inst{4}}
\institute{
  \inst{1} SMC-INFM \& Dipartimento di Fisica, Universit\`a ``La
Sapienza'' - P.le A. Moro 2, I-00185 Rome, Italy\\ 
  \inst{2} INFM \& Dipartimento di Ingegneria dell'Informazione,   
Seconda Universit\`a di Napoli, 
I-81031, Aversa, Italy\\
  \inst{3} NEST-INFM \& Dipartimento di Fisica "E. Amaldi",
Universit\`a di Roma Tre -- Via della Vasca Navale 84, I-00146 Roma,
Italy\\ 
  \inst{4} NEST-INFM \& Scuola Normale Superiore, I-56126 Pisa, Italy 
} 

\begin{document}

\maketitle

%\abstract{
\begin{abstract}
In this Letter we study the Kondo effect in an ultrasmall metallic grain, 
i.e. small enough to have a discrete energy-level spectrum,
by calculating the susceptibility
$\chi$ of the magnetic impurity.
Our quantum Monte Carlo simulations, and analytic solution of a simple
model, show that the behavior changes dramatically depending on
whether the number of electrons in the grain is even or odd.
We suggest that the measurements of $\chi$ 
provide an effective experimental way of probing the
grain's number parity.  
%\PACS{
%{75.75.+a}{Magnetic properties of nanostructures} \and
%{73.63.-b}{Electronic transport in mesoscopic or nanoscale materials
% and structures}
% } % end of PACS codes
% } % end of PACS codes
\end{abstract}
%
%\maketitle

Small metallic particles, typically of dimensions of few nanometers, 
show properties which are intermediate between atomic 
and bulk condensed matter systems. Because of the finite dimensions,
the quasi-continuous spectrum in the bulk splits into discrete energy
levels. When the average level spacing $\delta$ is comparable with 
the other energy scales of the problem interesting new effects appear.
For example the physical observables are parity dependent, i.e. 
differ if the number of electrons in the grain is even or odd. 
While  metallic particles have been investigated in ensembles 
of grains in the past (see~\cite{Perenboom} and references therein),
only recently spectroscopy of single grains~\cite{Black} has been 
achieved experimentally.
In the case of superconducting, ferromagnetic and/or antiferromagnetic systems, 
the effect 
of a finite level spacing is to enhance quantum fluctuations and 
consequently suppress the many-body condensation. 
As a result a  superconductor will not possess a 
fully developed gap~\cite{vonDelft}, the magnetization  of a 
ferromagnet will not be macroscopically large~\cite{Oreg},
although distinct features reminiscent of the macroscopic order can 
be detected even in nanosize systems.

Recently Thimm et al.~\cite{Thimm}, in their work on the so-called Kondo box,
studied the effect of magnetic impurities in a small metal particle. 
They found that the Kondo resonance is strongly affected when the mean 
level spacing in the grain becomes larger than the Kondo 
temperature~\cite{Thimm,Simon,Cornaglia}. The Kondo box shows the parity effect as well.
The Kondo effect~\cite{Hewson}  is a paradigm in correlated electron systems.
It originates from the interaction between the localized spin of a single
magnetic impurity and the free electrons of the metal. 
In bulk systems Kondo correlations manifest in an increase 
of the resistance for decreasing temperature $T$, above a
characteristic scale $T_K$, the Kondo temperature. If, instead, 
one looks at the magnetic properties of metals hosting Kondo 
impurities, one effect is the rapid saturation, for
decreasing $T<T_K$, of the impurity magnetic susceptibility $\chi$
(measuring the effect of an external magnetic field on the
impurity magnetization $M$)~\cite{WilsonRevModPhys}.
Mesoscopic physics is another arena in which the Kondo effect 
is playing a major role. By now, several years since the theoretical 
prediction~\cite{Glazman,Ng} a series of beautiful experiments have 
seen it in different devices (see the reviews~\cite{vanderWiel,Kouwenhoven}).
 In these cases the resonant level is realized by suitable fabricated 
nanostructure and the conduction electrons are provided by the metallic 
electrodes. 

In this Letter we are interested in the Kondo box, and we want to  
address the parity effect by means of the analysis of the impurity 
susceptibility $\chi(T)$, since 
$\chi(T)$ is very sensitive to the number of electrons in the system
as demonstrated in superconducting dots~\cite{DiLorenzo,Falci}.
Detection of parity effects in various thermodynamical quantities,
including the spin susceptibility, were measured for an ensemble 
of small, normal metallic grains~\cite{Volokitin96}. In similar 
fashion parity effects can be measured in an ensemble of small grains 
containing Kondo impurities. As it will be discussed in more details, 
also in this case, due to parity effects, grains with odd number of 
electron will give the dominant contribution to  $\chi(T)$.

The system is described by the Anderson model Hamiltonian
\begin{equation}\label{model}
H=\sum_{n,\sigma}(\epsilon_n-\mu  )
c^{\dagger}_{n,\sigma}c_{n,\sigma} + \sum_{\sigma}(\epsilon_d
-\mu)d^{\dagger}_{\sigma}d_{\sigma}
+Un_{d,\uparrow}n_{d,\downarrow}
+V\sum_{n,\sigma}(c^{\dagger}_{n,\sigma}d_{\sigma} +
c_{n,\sigma}d^{\dagger}_{\sigma})~,
\end{equation}
where $c^\dagger_{n,\sigma}$ ($c_{n,\sigma}$) is the 
rising (lowering) operator in the grain electron state with 
spin $\sigma=\pm 1/2$ (or $\uparrow$, $\downarrow$) and
energy $\epsilon_n=\epsilon_0+n\delta$, labeled by the
quantum number $n$
starting from the lowest level $\epsilon_0<0$ and
$d^{\dagger}_{\sigma}$ ($d_{\sigma}$) is the 
rising (lowering) operator in the impurity electron state with 
spin $\sigma$ and energy $\epsilon_d$ with the associated 
number operator for spin up (down) in the impurity state
$n_{d,\uparrow}\equiv d^{\dagger}_{\uparrow}d_{\uparrow}$
($n_{d,\downarrow}\equiv d^{\dagger}_{\downarrow}d_{\downarrow}$).
The Fermi level is taken as the zero of the  energy. 
The number of electrons $N$ in the grain is selected by fixing 
$\epsilon_0$, $\delta$, and the 
chemical potential to $\mu = 0$ for an odd number and $\mu=\delta /2$ 
for an even number, with 
%\begin{equation}
$(\epsilon_0-\mu)+(N-1)\delta/2=0$.
%\end{equation}
Only the levels up to $-\epsilon_0>0$ are relevant to the problem and
the density of grain electrons in the bandwidth 
$2|\epsilon_0|$ is assumed constant. The energy $U>0$ accounts for the 
on-site repulsion in the localized state.
The fourth term in Eq.(\ref{model}) describes the hopping 
of an electron from a grain level $\epsilon_n$ into
the impurity level $\epsilon_d$  and vice versa, with
a hybridization energy $V$. The level width $\Gamma$ scales as
$\Gamma =\pi V^2/\delta$ and remains finite for increasing $\delta$, 
because, as discussed in Ref.~\cite{Thimm}, both $\delta$ and $V^2$ scales 
as the inverse of the grain volume. This is consistent with the increase 
of the probability ($\sim V^2$) of finding an electron on the impurity 
site for decreasing grain volume. 

\begin{figure}
\onefigure[width=6.cm,height=5.2cm,angle=0]{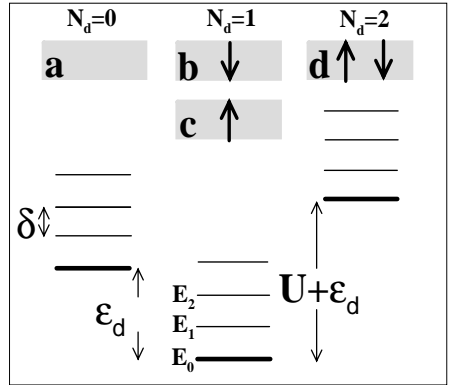}
\caption{%
Schematic representation of the energy levels for a system composed of 
a magnetic impurity
embedded in a nanoscopic metallic grain in the Anderson model with $V=0$.
The possible states for the impurity level are:
($a$) empty level ($N_d=0$) with $M=0$;
($b$) or ($c$) singly-occupied level 
($N_d=1$) with 
$M=-1/2$ (or spin $\downarrow$)
or
$M=1/2$ (or spin $\uparrow$);
($d$) doubly occupied 
level ($N_d=2$) with electrons with anti-parallel spins
($\uparrow\downarrow$) and $M=0$.
The corresponding energies 
are represented, in the bottom part, by bold lines for 
the states with the electrons filling the grain levels
$\epsilon_n$ up to the Fermi level and the impurity level empty ($a$), 
the impurity level singly-occupied 
with energy gain $\epsilon_d<0$ ($b$ or $c$), 
the impurity level doubly-occupied 
by electrons repelling each other with 
energy $U>0$ ($d$).
Thin lines represents the {\it excited} states with one or more 
grain electrons promoted to the first empty state above the Fermi
level. 
}
\label{sketch}
\end{figure}
In order to put in context our analysis, let us recall some basic
concepts concerning the physics associated with the Anderson model.
The impurity site  may be in one of the
four different states: empty, singly occupied with up (down) spin,
and doubly occupied. The magnetic properties of the impurity depend on the
occupation probability of each state.
As discussed in  Ref.~\cite{Krishna}, depending on $T$,  
one has  different regimes. To illustrate how these different regimes
develop,  it is useful to consider first  the case with no 
hybridization ($V=0$) and $\epsilon_d <0$ (Fig.~\ref{sketch}).
For $k_BT\gg U+\epsilon_d$ (where $k_B$ is the Boltzmann constant),
the four impurity states 
have the same probability to occur. 
Therefore the probability to have two electrons in the impurity level
is $P_2=1/4$, the average number of electrons in the impurity level is
$N_d=1$ and the average of the squared value of the total spin in the
impurity level is $M^2=1/8$. 
The impurity susceptibility as a function of $T$ is, in this case, 
$\chi_0(T)=T_K/(2T)$
(hereafter we use the dimensionless susceptibility 
$\chi_0\equiv 4\chi k_BT_K/(g\mu_B)^2$,
where  $g$ is the 
Land\'e
factor and $\mu_B$ is the Bohr
magneton), as in the case of a 
{\it free orbital}. 
For $U+\epsilon_d>k_BT>-\epsilon_d$ the double occupancy is
excluded,
resulting in $P_2\simeq 0$, $N_d\simeq
2/3$, $M^2\simeq 1/6$ and $\chi_0(T)\simeq 2T_K/(3T)$.
%({\it valence-fluctuation} regime).
For $k_BT<-\epsilon_d$ only the  
single occupied states have a finite probability to occur, 
with $P_2\simeq 0$, $N_d\simeq 1$,
$M^2\simeq 1/4$ and $\chi_0(T)\simeq T_K/T$. 
The impurity properties in this case are like those of a 
{\it local (magnetic) moment}, i.e. of an isolated electron able to
flip its spin $\sigma$. 

If $V>0$, the electrons can
tunnel between the impurity level and the grain levels. 
The virtual transitions to doubly occupied and empty impurity states lead
to an effective antiferromagnetic interaction 
$J_{eff}$ between the grain electron spin density  and the impurity 
spin. 

For a large grain's size ($\delta\rightarrow 0$), 
in first approximation (to the lowest-order of the 
perturbation theory~\cite{Hewson}), is $J_{eff}\sim \Gamma/U$.
The higher orders can be taken into account by the renormalization group
(RG) approach via an energy-dependent interaction $J_{eff}(E)$
\cite{Hewson,Krishna}.
The RG procedure shows that, at low $T$, $J_{eff}(E)$ increases 
({\em strong coupling} regime)~\cite{Anderson}. 
Below $T_K$,
the impurity spin becomes effectively bound into a singlet (total spin zero) with
the grain electrons' spin.
As a consequence in
the strong-coupling regime, 
the impurity susceptibility saturates, for $T\rightarrow 0$, to
$
\chi_0(\delta=0,T=0)\simeq 0.4128 ~ .
$

\begin{figure}
\begin{center}
\onefigure[width=7cm,height=7cm,angle=0]{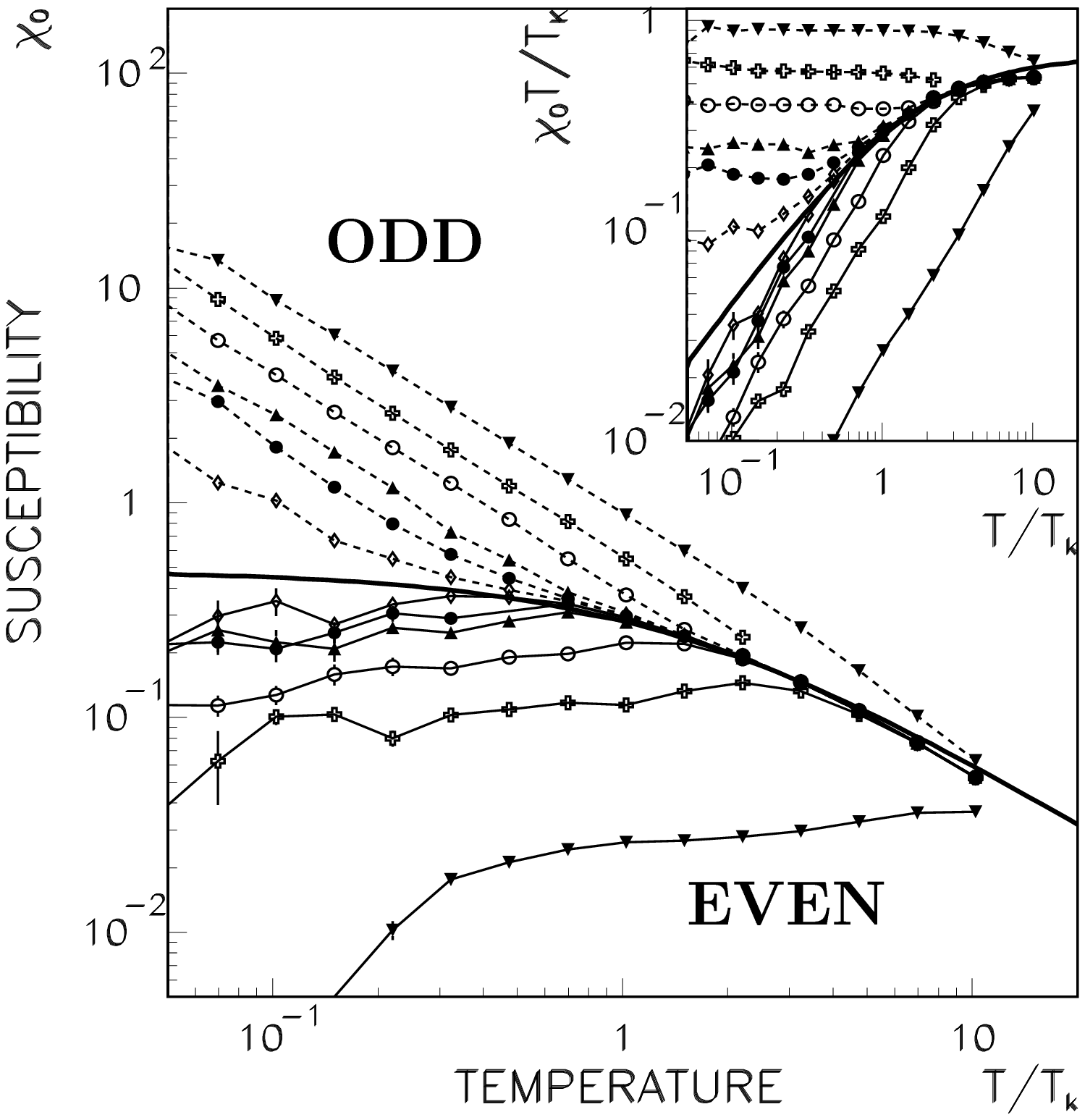}
\end{center}
\caption{%
The parity effect on the magnetic susceptibility.
The (dimensionless) susceptibility $\chi_0$,
as function of the (dimensionless) temperature $T/T_K$,
for a magnetic impurity embedded in a non-magnetic metallic grain,
shows a clear parity effect when the level spacing $\delta$ is large:
dotted lines correspond to odd number $N$ of electrons in the system,
solid lines to even $N$.
We show the 
quantum Monte Carlo calculations for $L=300$ 
for the symmetric Anderson model with 
$\Gamma/(\pi k_BT_K)\simeq 1.62\ll U/(k_BT_K)\simeq 20.45$ and 
$|\epsilon_0|=2556.25 k_B T_K$, for
$\delta/(k_BT_K)\simeq 0.01$ (bold line), recovering the
$\delta\rightarrow 0$ limit, and, starting from the bold line and
going outward, for 
$\delta/(k_BT_K)\simeq 2.04$ (rhombus) 
3.37 (full circles),
5.11 (up triangles), 
10.22 (open circles),
25.56 (crosses),  
and 255.57 (down triangles).
The corresponding values of $N$ range from 2508 (solid line with rhombus) 
to 21 (dotted line with down triangles).
Where not shown, errors are smaller than symbols.
Inset: the same results times $T/T_K$ are shown to emphasize the local
moment regime for $k_BT\simeq U/2$.
}
\label{chi}
\begin{center}
\twofigures[width=7cm,height=7cm,angle=0]{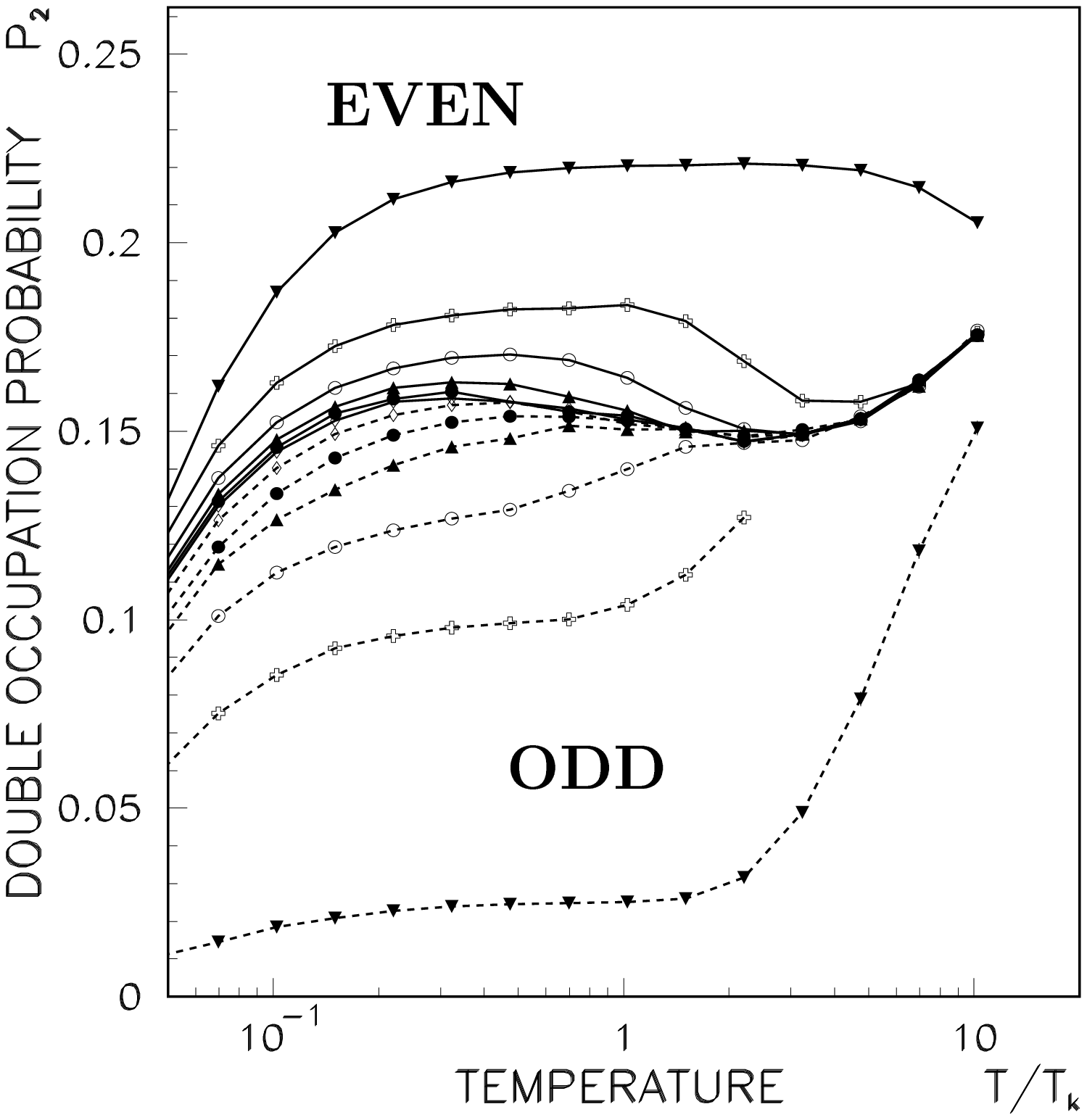}{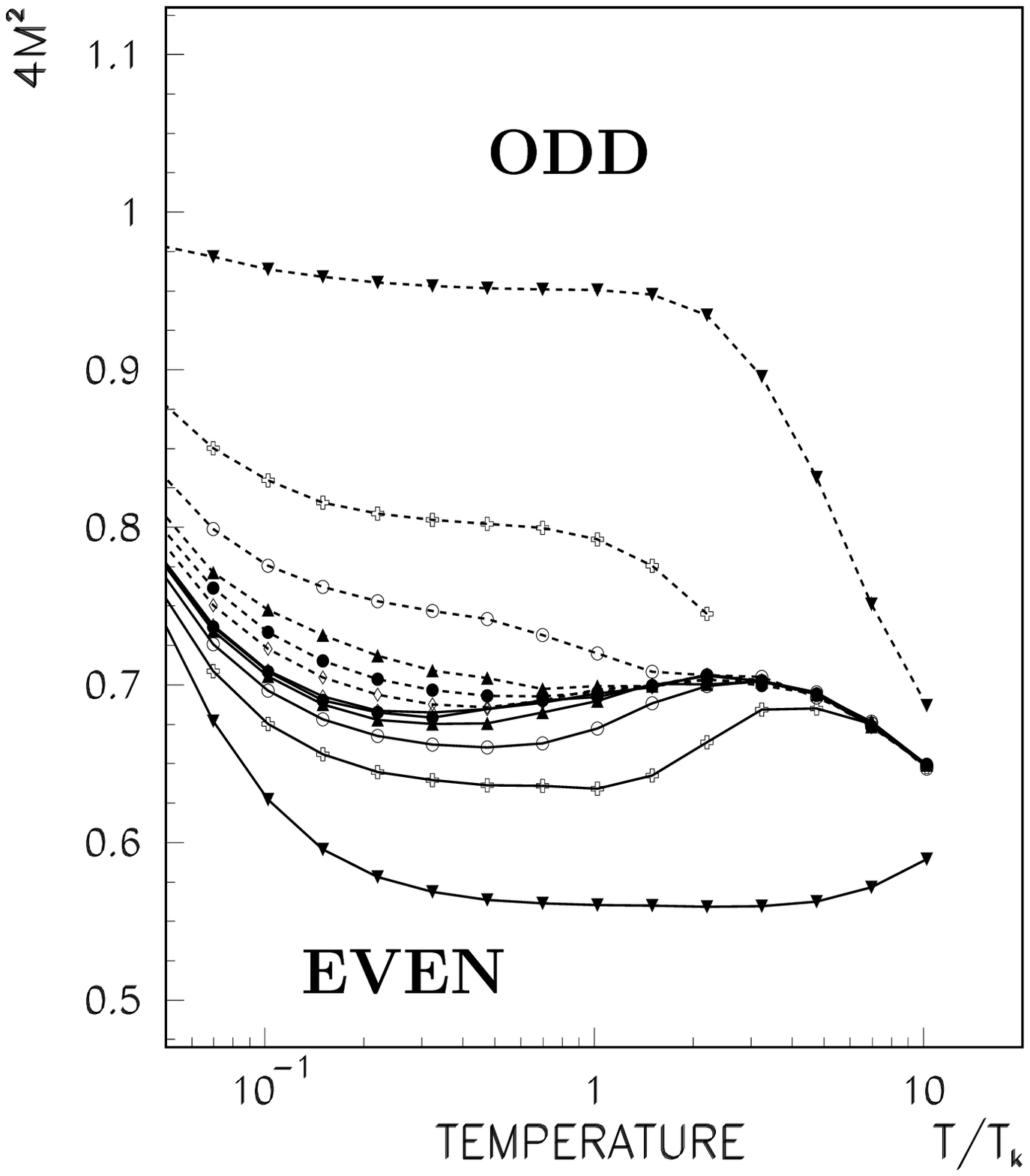}
\end{center}
\caption{%
The parity effect on the probability $P_2$ of finding two electrons
localized on the impurity.
Symbols and lines are like those in Fig.\protect~\ref{chi}.
The calculations for $\delta\simeq 0.01 k_BT_K$ (bold line) show
a non-monotonic behavior with a maximum for $T<T_K$. 
Dotted lines are for odd $N$; solid lines for even $N$.
The calculations are for $L=200$.
Comparison with calculations for $L=300$ do not show any relevant 
finite-size effect for $\delta>1.5 k_BT_K$.%
}
\label{p2}
\caption{The parity effect on the squared impurity magnetization
 $M^2$. Symbols and lines are like those 
 in Fig.~\protect\ref{p2}. Since $N_d=4M^2+2P_2$, all our calculations have
 $N_d\simeq 1$.}
\label{4m2}
\end{figure}

In ultrasmall grains, a finite 
$\delta$ introduces a further low-energy scale beside $T_K$. In order 
to describe the effect of finite level spacing on the Kondo resonance
we use a quantum Monte Carlo approach proposed by Hirsh and Fye
\cite{Hirsh86,Georges96}. According to this method, after integrating 
out the grain level degrees of freedom, the resulting impurity problem 
is mapped into a chain of two-states auxiliary variables 
(fictitious Ising spins),
independent on each other, but interacting
with an effective magnetic field, proportional to the impurity
magnetization $M$. 
The length $L$ of the chain is proportional to the time-evolution of
the system.
The method is exact in the limit of an infinite chain of Ising spins and 
gives results reliable within an approximation $\sim (TL)^{-2}$.
To study the properties of the impurity spin 
we calculate
$\chi$, $P_2$, $N_d$ and $M^2$ as a function of $T$,
and $L$ (verifying, as a check,
the exact relation $4M^2=N_d-2P_2$ valid at any $T$). 
We consider the {\it symmetric} case (i.e. with $\epsilon_d=-U/2$).
%where there is no valence-fluctuation regime.
We estimate $T_K$ for $\delta\rightarrow 0$, by fitting our calculations  
with the low-$T$ approximation~\protect\cite{Hewson}
\begin{equation}
\frac{4\chi k_B}{(g\mu_B)^2}= \frac{0.68}{T+1.4142T_K}
\label{lowT}
\end{equation}
where $T_K$ is the only free parameter.
In this limit  our results (Fig.~\ref{chi})
recover, at $k_BT\simeq -\epsilon_d=U/2$, 
$\chi_0 T/T_K\simeq 1$ 
(local moment regime) and, for $T\leq T_K$, 
$\chi_0=0.4128$ (strong coupling regime).
For $\delta\geq 2 k_BT_K$,
$\chi_0$ (Fig.~\ref{chi}),
$P_2$ (Fig.~\ref{p2}) and $M^2$ (Fig.~\ref{4m2})
reveal a clear parity effect.
The main features are the following. 

{\it For an odd number $N$ of electrons} in the system 
(dotted lines in Fig.~\ref{chi}-\ref{4m2}),
i) 
the local moment regime extends 
down to the lowest $T$, with 
$\chi_0\sim 1/T$ (Curie law);
ii)
$\chi_0 T/T_K$ saturates to a value
that increases with $\delta$, approaching the value $1$ expected for
$\Gamma=0$, and
iii)
for $T<T^*$ and increasing $\delta$, 
$P_2$ decreases toward 0 as expected in the local moment regime.

{\it For an even $N$} 
(solid lines in Fig.~\ref{chi}-\ref{4m2}),
iv) the Kondo effect appears
to be enhanced and
$\chi$ saturates to a value that decreases with increasing
$\delta$. 
This can be seen as an increase of the effective Kondo
temperature $T_K(\delta)$ on the base of the Eq.(\ref{lowT}).
v)
For %even $N$, 
$T<T^*$ and increasing $\delta$, 
$P_2$ increases toward $1/4$ as in the free orbital regime.

Finally,
vi)
for $k_BT<V$, with the hybridization energy $V\sim\delta^{1/2}$,
the non-monotonic behavior of $P_2$ is depressed when $N$ is {\it odd} and 
emphasized when $N$ is {\it even};
vii) 
for any $N$,
the onset of the parity effect increases with $\delta$ and, 
empirically, the onset is at $k_BT^* \simeq \delta/5$ for all the
cases considered. 
The proportionality of the onset 
$T^*$ to $\delta$, %as in vii), 
is expected, because, for
$k_BT>\delta$, the approximation of the grain levels with a band is
valid. 

To understand the {\it odd} case, consider the limit
$\delta\rightarrow\infty$, equivalent to a system with 
an impurity level at energy $\epsilon_d<0$ and a 
doubly occupied grain level at (zero) Fermi energy. 
The only possible process in this case is a transition of a grain
electron to the impurity level, but for $k_BT<U$ this transition is
suppressed, because the double occupancy in the impurity level costs
an energy $U$. 
Therefore, the impurity electron behaves as an isolated magnetic
moment, with $P_2\simeq 0$ and 
following the Curie law $\chi_0\sim 1/T$ (Fig.~\ref{chi}), as in i).
If $\delta$ is finite, also the transition of the impurity electron
above the Fermi level takes place, but requires an energy
$|\epsilon_d|+\delta$ and occurs with probability $\sim
\exp[-\delta/(k_BT)]$.
Therefore, for large $\delta$ and low $T$, this
transition is suppressed and $\chi_0 T/T_K\rightarrow 1$ (inset
Fig.~\ref{chi}), as in the local moment regime, and $P_2\rightarrow
0$ (Fig.~\ref{p2}), as in iii).
For decreasing $\delta$, the system enters the local moment regime
only at $T<T^*\sim \delta$, with $\chi_0 T/T_K$
saturating at the value reached at $T^*$. Therefore,
the lower $\delta$, the lower the asymptotic value (inset Fig.~\ref{chi}), 
as in ii).
%to an effective Kondo temperature $T_K(\delta)\ll T_K(\delta=0)$.

\begin{figure}
\onefigure[width=7cm,height=7cm,angle=0]{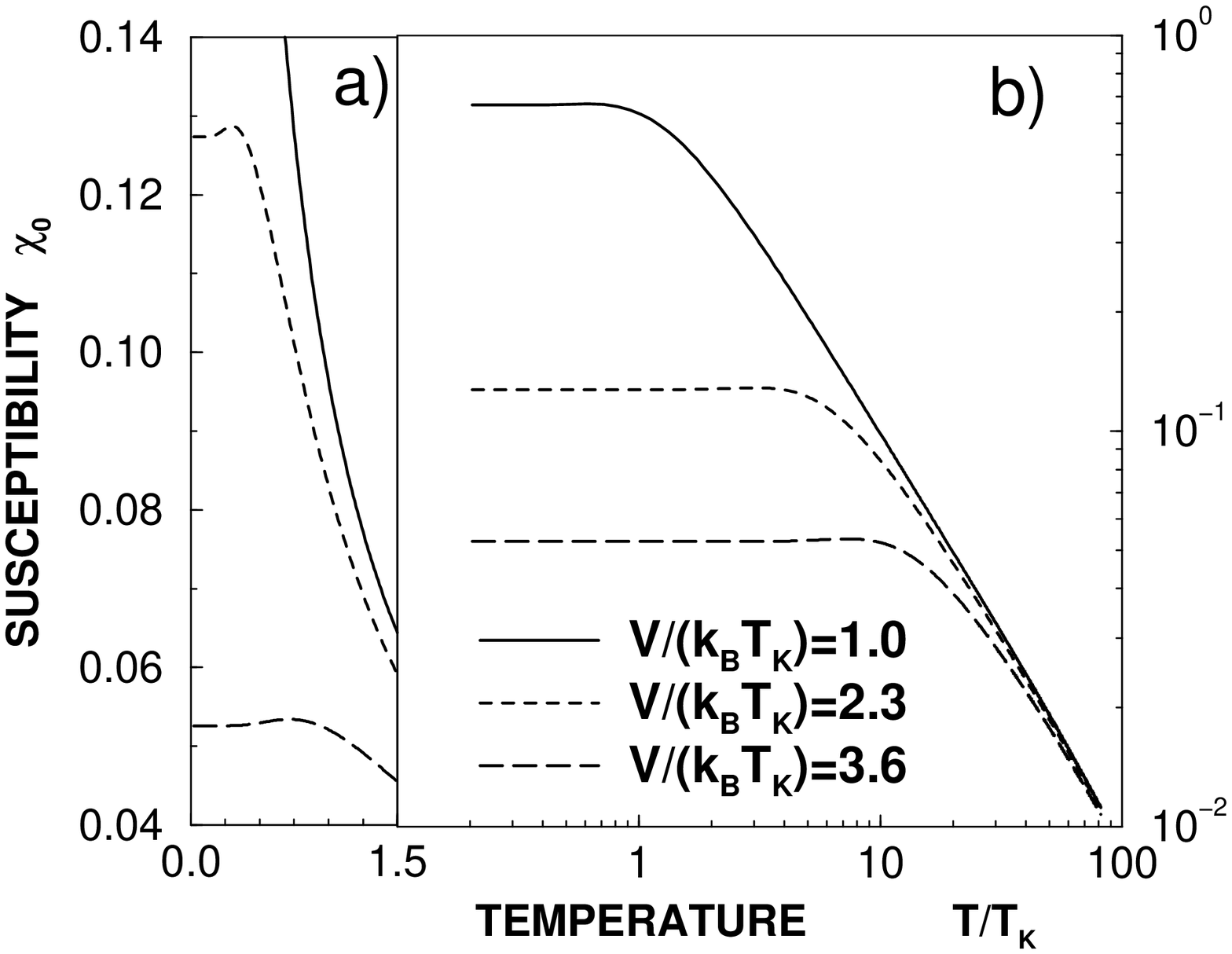}
\caption{%
The exact susceptibility $\chi_0$ 
for {\it even} $N$ and $\delta\rightarrow\infty$, as a function of $T/T_K$.
The calculations for the symmetric Anderson model with $U=20.44
k_BT_K$ and $V/(k_BT_K)=1.0$ (solid line), 2.3 (dashed line) and 3.6 
(long-dashed line), are shown in linear scale, in a) to emphasize the
maxima of $\chi_0$, and in double logarithmic scale, in b), to emphasize
the high-$T$ behavior $\chi_0\sim 1/T$ (the linear part in b).%
}
\label{analytic_chi}
\end{figure}

In the {\em even} case, 
when there is only one electron in the Fermi level, 
transitions between the Fermi level and the impurity level are allowed and this
results in a 
strong hybridization between the two levels. 
%-----
%for increasing $\delta$ and $T<T^*$, 
%$P_2$ increases (Fig.~\ref{p2}) 
%and $\chi_0$ saturates to a decreasing $\chi_0(T=0)$ (Fig.~\ref{chi}), 
%that can be interpreted as an
%increase of an effective $T_K(\delta)$ [Eq.(~\ref{chi_0})].
%
We perform in this case an
%recover this behavior by means of the
exact calculation of $\chi_0(T)$ for $\delta\rightarrow \infty$ 
(Fig.~\ref{analytic_chi}).
%%RR
In this limit, only  the grain level at the
Fermi energy couples with the impurity,
resulting in %and one is left with 
an interacting two-level system.
For very large, but finite $\delta$, this two-level system
is still a good approximation since
the grain levels away from the Fermi level 
can be taken into account in perturbation theory up to second order.
Their main effect is to renormalize the impurity parameters.
In particular,
the energy of the impurity level is renormalized to a %pushed up in 
higher energy, closer to the Fermi level. 
This is just what is left of the logarithmic renormalization,
familiar in the bulk system.
%%RR
The two-level system has 
six possible different many-body two-electron states:
$(1)$   with one electron with spin $\uparrow$ in the Fermi level and
one with spin $\downarrow$ in the impurity 
$|1\rangle=c^{\dagger}_{\uparrow}d^{\dagger}_{\downarrow}|0\rangle$;
$(2)$  with inverted spins with respect to the first state
$|2\rangle=c^{\dagger}_{\downarrow}d^{\dagger}_{\uparrow}|0\rangle$;
$(3)$ with two electrons with opposite spins in the Fermi level
$|3\rangle=c^{\dagger}_{\uparrow}c^{\dagger}_{\downarrow}|0\rangle$;
$(4)$  with two electrons with opposite spins in the impurity level
$|4\rangle=d^{\dagger}_{\uparrow}d^{\dagger}_{\downarrow}|0\rangle$;
$(5)$   with one electron with spin $\uparrow$ in the Fermi level and
one with spin $\uparrow$ in the impurity
$|5\rangle=c^{\dagger}_{\uparrow}d^{\dagger}_{\uparrow}|0\rangle$, and
$(6)$  with inverted spins with respect to the fifth state
$|6\rangle=c^{\dagger}_{\downarrow}d^{\dagger}_{\downarrow}|0\rangle$.
These six states form three singlet ($M=0$) and one triplet ($M=0,\pm
1$) states. 
The singlet states are given by $(|1\rangle-|2\rangle)/\sqrt{2}$,
$|3\rangle$, and $|4\rangle$. 
The triplet components are instead given by   $(|1\rangle+|2\rangle)/\sqrt{2}$,
$|5\rangle$, and $|6\rangle$.
To probe the impurity susceptibility we couple exclusively the impurity spin
with an external magnetic field $h$.
The model, in this case, can be described by a diagonal block matrix,
with a $4 \times 4$ block $H_a$, for the states from (1) to (4), and a 
$2 \times 2$ block $H_b$, for the states (5) and (6), with
\begin{equation}
H_a =	\left(
	\begin{array}
	{c c c c}
	\epsilon_d +h	& 0		& V	& V		\\
	0		&\epsilon_d -h	&-V	&-V		\\
	V		&-V		&0	&0		\\
	V		&-V		&0	&2\epsilon_d+U  \\
	\end{array}
	\right)
~~~\mbox{    and    }~~~
H_b =\left(\begin{array}{c c}
\epsilon_d -h& 0\\
0&\epsilon_d +h
\end{array}
\right) ~,
\label{4by4}
\end{equation}
%and
%\begin{equation}
%\label{2by2}
%\end{equation}
where the diagonal elements are the energies in each of the six
states, and the square of an off-diagonal element is the
transition probability between the corresponding states.
By computing the energies $E_i$ corresponding to the {\it eigenstate}
of these two matrices, we calculate the impurity magnetization 
$m(T,h)=\sum_i M_i \exp[E_i/(k_BT)]$, where $M_i$ is the impurity
magnetization of the {\it eigenstate} $i$, and
$\chi(T)\equiv (\partial m/\partial h)_{h=0}$.

We find that $\chi\sim 1/T$ for $k_BT>V$ (Fig.~\ref{analytic_chi}b).
Indeed, for $k_BT>V$ the states $|1\rangle$, $|2\rangle$, $|5\rangle$
and $|6\rangle$  
have the same energy (for $h=0$), while states $|3\rangle$ and $|4\rangle$ have
higher energy, therefore the impurity level is singly occupied
(local moment regime). 
For $k_BT<V$, we find $\chi$ saturating to a constant value, 
decreasing with increasing $V$ (Fig.~\ref{analytic_chi}).
Indeed, as a consequence of the hybridization $V$, the ground state is
given by the lowest-energy singlet, that has an energy difference
with the triplet states larger than $V$. Therefore, $\chi$ is constant
and the asymptotic value is approximately given by the Curie law at 
$k_BT\simeq V$. Therefore, the larger $V$ (or $\delta$), the smaller
the saturation value of $\chi$ for $T\rightarrow 0$, as in iv).
All these results are indeed consistent with our quantum Monte Carlo
simulations (Fig.~\ref{chi}), by  keeping in mind that $V$ scales as $\delta^{1/2}$ for
fixed transition rate $\Gamma$. 

Moreover, 
the actual diagonalization shows that the transition from the
local moment to the saturation regime occurs via 
a non-monotonic behavior (with a maximum depending on the
hybridization $V$). This result is not clearly detectable in our
quantum Monte Carlo calculations for $\chi$ (Fig.~\ref{chi}), but is consistent with
the non-monotonic behavior of $P_2$ (Fig.~\ref{p2}), more evident in the even case, that
shows that the local moment regime extends to lower $T$, as in vi).
In particular, to understand in the even case the behavior of $P_2$
for finite decreasing $\delta$, we should consider more grain levels
in our exact solution.
For example, by including one grain level 
below the
Fermi energy and one above, the many-body states are 
four-electron states,
due to three grain levels and one impurity 
level.
As noted before, the hybridization favors a singlet ground state.
In this case the weight of the singlet state with doubly-occupied
impurity level is decreased, because the singlet state manifold has a
larger number of components. As a consequence, the probability $P_2$
decreases with decreasing $\delta$ (Fig.~\ref{p2}), as in v).

In summary, we have shown that, a magnetic impurity embedded in a small
metallic grain, has a striking different behavior depending on the
parity of the electron number in the grain.
In particular,
the Kondo effect is strongly enhanced if the number of electrons is
even, while it is strongly depressed if the number is odd.
Therefore,
the measurement of $\chi$ of a magnetic
impurity embedded in an ultrasmall metallic grain is an effective
way of probing the number parity of the electrons in the grain.

\acknowledgments
We acknowledge fruitful discussions with G. Falci, A. Mastellone, and A. Tagliacozzo.
This work was supported by INFM under the PRA-project
"Quantum Transport in Mesoscopic Devices" and EU by Grant RTN 1-1999-00406,
RTN2-2001-00440, andHPRN-CT-2002-00144.

\end{document}